\documentclass[usegraphicx,useAMS,usenatbib]{mn2e}

\usepackage{color}

\title{A possible link among pulsar timing noise  intermittency  and  hidden ultra-compact binaries}

\author[Gong $\&$ Li]
    {B.P. Gong$^{1}$\thanks{E-mail: bpgong@mail.hust.edu.cn} , Y.P. Li$^{2}$\thanks{E-mail: liyp@stu.xmu.edu.cn}\\
$^1$Department of Physics, Huazhong University of Science and
Technology, Wuhan 430074, China,  \\
$^2$Department of Astronomy, Xiamen University, Xiamen, 361005, China \\
}
\begin{document}

\date{Accepted, Received ;}

\pagerange{\pageref{firstpage}--\pageref{lastpage}} \pubyear{2012}

\maketitle

\label{firstpage}









\begin{abstract}
{The  quasi-periodic feature of 1-10 years exhibited in pulsar timing noise has not been well understood since 1980. The recently demonstrated  correlation between timing noise  and  variation of pulse profile motivates us to further investigate its origins.
We suggest that the quasi-periodicity feature of  timing noise, with rapid oscillations lying on lower frequency structure,
comes from  the geodetic precession of an unseen binary system,
which  induces additional motion of the pulsar spin axis. The resultant change of  azimuth and latitude
at which the observer's line of sight crosses the emission beam is responsible for the variation of  timing noise and pulse profile respectively.
The first numerical simulation to both timing noise and
pulse profile variation are thus performed, from which
the orbital periods of these pulsars are of 1-35 minutes. Considering the existence of the ultracompact binary white dwarf of  orbital period of  5.4 minute, HM Cancri, such orbital periods to pulsar binaries are not strange.
The  change of latitude of the  magnetic moment exceeding the range of emission beam of a pulsar  results in  the intermittency, which explains the behavior of PSR B1931+24.
Therefore, it  provides not only a mechanism of the quasi-periodic feature displayed on some ``singular" pulsars, like timing noise,  variation of pulse profile and  intermittency; but also a new  approach of searching ultra-compact binaries, possibly pulsar-black hole binary systems.
}
\end{abstract}


\section{Introduction}


Pulsar timing noise is the discrepancy between  pulsar arrival times at an observatory and times predicted from a spin down model, containing the rotational frequency and its derivative.
Such low-frequency structures previously observed in pulsar data sets have been explained by random processes~\citep{Cordes80,Lyne00}, unmodelled planetary companions~\citep{Cordes93} or free-precession~\citep{Stairs00}. However, these models predict either random or exact periodic timing delay, which cannot well account for the quasi-periodic feature  revealed, i.e., in the large investigation of pulsar rotation properties~\citep{Hobbs10}.

Lyne et al. (2010) reported the correlation of timing noise with changes in the
pulse shape in six pulsars. And the link among mode changing~\citep{Wang07}, nulling,
intermittency, pulse-shape variability, and timing noise is  addressed.  These
phenomena are attributed to the change in the pulsar's magnetosphere~\citep{Lyne10}.


This paper shows that  the  geodetic precession induced motion of pulsar spin axis can change both the latitude and azimuth of the magnetic moment of the pulsar, which mimics the  magnetosphere change, and hence  well account for the quasi-periodic structure and its correlated phenomena.


In the absence of spin precession, the magnetic moment of the pulsar, $\mu$, with a misalignment angle, $\alpha$, relative to  the spin axis,
rotates around a fixed axis~\citep{Rad69}. The observed pulsed emission occurs when the pulsar beam sweeps past the line of sight, in the case of dipole magnetic field, as shown in Fig.~1.

If a pulsar is in orbit with a rotating companion star, then the two spinning tops are in the gravitation field of each other, and both the spin axes  will precess around the common orbital angular momentum vector. This is the so-called  geodetic precession.
The  precession of a pulsar spin axis  will cause
additional motion of the magnetic moment, $\mu$, owing to a fixed, $\alpha$, the misalignment angle between the spin axis and $\mu$. Consequently, both the azimuth  and latitude at which the observer's line of sight crosses the beam change, which in turn leads to quasi-periodic timing delay and variation in pulse profile respectively.
However, these pulsars with significant timing noise are usually identified as isolated pulsars instead of binary ones.
This problem can be explained by three reasons.

Firstly, the current limit on binary search  is 90 minutes for radio pulsars, and 10 minutes for  X-ray pulsars. In other words, pulsars with orbital period shorter than these thresholds  have not been detected.
But from the stand of evolution of binary pulsars, it is difficult to understand why binary  pulsars with orbital period less than 90 minutes or 10 minutes cannot exist. Obviously, the limit is of technique rather than true existence. Therefore, it is possible that some of these isolated pulsars with significant timing noise are binary pulsars beyond the threshold of binary detection.

Secondly,  e.g., PSR 1828-11 has  several 8-min continuous time series observed by Parkes telescope (R.N. Manchester, private communication), but none of them  cover a full proposed orbital period of $\sim$20 min. Thus, orbital modulation on the timing of PSR 1828-11 has never been observed in over a complete orbital period, although it  has been observed for years.


Thirdly, the short-term effect of a binary system, i.e., the time delay corresponding to the propagation time of pulsar emission during its orbital motion, the R\"omer delay,   depends on the amplitude of  semi-major axis of a binary, $a$. Apparently a small $a$ leads to low signal to noise ratio in time of arrival. Moreover,  ultra-compact binaries  correspond to rapid orbital motion which prevents from having  good quality profile (to discern orbital modulation) in such short orbital bins.


Whereas, the geodetic precession induced additional time delay is a long-term effect (usually lasts a few orders of magnitude longer than that of the short-term effect), the amplitude of which can be large in the case of small $a$. As a result, the orbital effect of binary pulsars with small $a$ is difficult to detect. In contrast, the corresponding long-term effect is still  easy to find. Therefore, such pulsars could be identified  as  singular pulsars with strong timing noise~\citep{Gong05,Gong06}.

How the geodetic precession invokes the quasi-periodic change of timing noise and pulse profile ?
In the gravitational two-body system, the geodetic precession makes one body precess in the gravitational field of the other,  which influences pulsar timing in two ways.

The first one is the spin vector of one body precesses around the orbital angular momentum vector,  the precession velocity of which is given by the two-body equation~\citep{BO75},
\begin{equation}\label{Omega}
 \dot{\mathbf \Omega}_{1}= \frac {L
(4+3m_{2}/m_{1})}{2r^{3}}\hat{\mathbf L}= T_{\odot}^{2/3}(\frac{2\pi}{P_{\rm b}^{\ast}})^{3/5}\times\frac{M_{\rm r}(4+3M_{\rm r})}{2(1+M_{\rm r})^{4/3}}\hat{\mathbf L} \\
\end{equation}
where $L$ and $\hat{\mathbf L}$, are magnitude and
unit vector of the orbital angular momentum, $m_1$ and $m_2$ are masses of the two bodies, and $r$ the separation between them;  $M_{\rm r}\equiv m_{1}/m_{2}$, $T_{\odot}\equiv GM_{\odot}/c^{3}=4.925490947\mu$s and $P_{\rm b}^{\ast}\equiv P_{\rm b}(1-e^2)^{\frac{3}{5}}/(m^{\ast}_{1})^{\frac{2}{5}}$ (where $m^{\ast}_{1}\equiv m_{1}/M_{\odot}$, $P_b$ and $e$ are orbital period and eccentricity respectively).
Notice that in  Eq.($\ref{Omega}$) the expression without $T_{\odot}$ is the original two-body equation~\citep{BO75}, and the one with $T_{\odot}$ is rewritten for the convenience of numerical calculation. Here after the observed pulsar is labeled by subscript 1, and its companion star by 2.

\begin{figure}
\begin{center}
\includegraphics[width=0.5\textwidth]{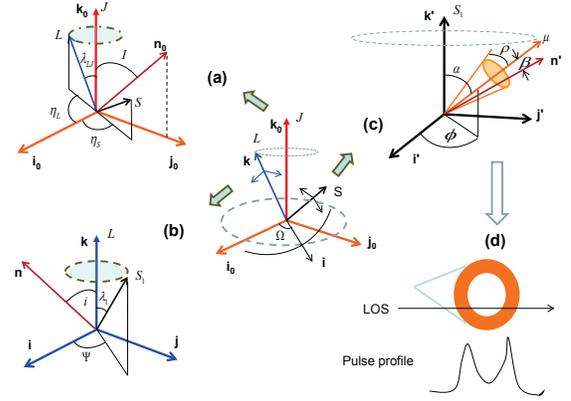}
\end{center}
\caption{\small Definition of the angles and
three coordinate systems. In the case of $\lambda_{LJ} \ll 1$, which can  always be satisfied in our calculation, the approximation, $\Psi = \eta_{1}-\Omega$ and $\lambda_{1L} = \lambda_{1}$ are  adopted,  where $\lambda_{1L}$ and $\lambda_{1}$ are the misalignment angle between $S_{1}$ and $L$, as well as  $S_{1}$ and $J$ respectively. \label{noise1}}
\end{figure}

As shown in  Fig.($\ref{noise1}$), the geodetic precession  can change the orientation of  spin axis, and thus affects
the azimuth  and the latitude at which the observer's line of sight crosses the beam. Such an additional motion of a pulsar spin axis
can be analyzed equivalently by assuming  a constant pulsar spin angular momentum vector, but a varying line of sight (LOS), in a reference frame defined by the spin angular momentum vector of the pulsar, $\mathbf{S}_1$,  ($\mathbf{i}^{\prime}$, $\mathbf{j}^{\prime}$, $\mathbf{k}^{\prime}$).

Another useful reference frame is defined by the angular momentum vector, $\mathbf{L}$,  ($\mathbf{i}$, $\mathbf{j}$, $\mathbf{k}$), in which the direction toward the observer, $\mathbf{n}$, is in the plane defined by the unit vectors,  $\mathbf{i}$ and $\mathbf{k}$,
$\mathbf{n}= (\sin \textit{i}, 0 , \cos \textit{i})$.
Transforming from the reference frame defined by $\mathbf{L}$, to the reference frame defined by $\mathbf{S}_1$, the LOS is changed from unit vector, $\mathbf{n}$ to $\mathbf{n}^{\prime}$, which is read~\citep{Konacki03},
\begin{equation}\label{ntran}
\mathbf{n}^{\prime}=R(-\lambda)R(\Psi)\mathbf{n} ,
\end{equation}
where $\mathbf{n}$ first rotates around the axis $\mathbf{k}$ by the angle, $\Psi$,  then around the axis $\mathbf{j}$  by angle, $-\lambda$.

A pulse is identified when the magnetic moment, $\mu$, crosses the plane defined by the vectors,
$\mathbf{n}^{\prime}$, and $\mathbf{S}_1$ and, at the same time, the angle between
$\mathbf{n}^{\prime}$, and $\mu$ does not exceed the opening semi-angle $\rho$, as shown in Fig.~1.
Obviously, the direction toward the observer $\mathbf{n}^{\prime}$ changes with the precession phase $\Psi$,
which can make a pulse arrive advanced or retarded than that of a fixed $\mathbf{n}^{\prime}$.
This effect can be quantified by the azimuthal angle of $\mathbf{n}^{\prime}$~\citep{Konacki03},
\begin{equation}\label{nyx}
\tan \Theta(t)=n^{\prime}_y(t)/n^{\prime}_x(t)
\end{equation}
Apparently, a two-body system with a single spin precessing as Eq.($\ref{Omega}$) will result in the timing effect of Eq.($\ref{nyx}$),  which predicts a  periodic timing effect similar to the free precession.

Here is the second way that the geodetic precession influences  pulsar timing.
 A two-body system with spin can cause an additional perturbation to the classical two-body system without spin, and hence, all six orbital elements of a binary vary, including  $\omega$ and $\Omega$,  the precession of periastron and the longitude of precession of the orbital plane respectively~\citep{SB76}. However, the contribution of $\omega$ and $\Omega$ to R\"omer delay depends on the projected semi-major axis, $x$~\citep{Lai95}. On the contrary, the precession induced long-term azimuthal time delay, as shown in Eq.($\ref{ntran}$) and Eq.($\ref{nyx}$), is not directly related with $x$, which can be significant in the case of small $x$.

The arrangement of this paper is as follows. Section 2 shows the long-term additional time delay and change of pulse width caused by $\Omega$, the precession speed of the orbital angular momentum vector, $\mathbf{L}$,  around the total angular momentum,  $\mathbf{J}$. Section 3 applies the model to three pulsars,  fitting their quasi-periodicity timing noise,  and simulating the corresponding variation of pulsar profile width and intermittency for the first time. The discussion and conclusion of  results are presented in the final Section.



\section{Long-term binary effect on pulsar timing}

When $\dot{\Omega}$, the precession of the orbital angular momentum vector, $\mathbf{L}$,  around the total angular momentum,  $\mathbf{J}$, which has been ignored by Eq.($\ref{ntran}$) and Eq.($\ref{nyx}$),  is considered,  the LOS should be transformed as,
\begin{equation}\label{nyxprime}
\mathbf{n}^{\prime}=R(-\lambda_{1})R(\Psi)R(\lambda_{LJ})R(\Omega)\mathbf{n}_0 ,
\end{equation}
where  $\lambda_{LJ}$ is the misalignment angle between  $\mathbf{L}$ and $\mathbf{J}$, and $\mathbf{n}_0$=$(0, \sin I, \cos I)$ is the unit vector of LOS defined in the reference frame  ($\mathbf{i}_0$, $\mathbf{j}_0$, $\mathbf{k}_0$) with $\mathbf{k}_0$ aligned with $\mathbf{J}$, as shown in Fig.~1. Thus, putting the components $n^{\prime}_x(t) $ and $n^{\prime}_y(t)$ obtained from  Eq.($\ref{nyxprime}$) into Eq.($\ref{nyx}$), the time delay including to the precession of orbital plane can be obtained.

Note that Eq.($\ref{nyxprime}$) has one more frame transformation than that of Eq.($\ref{ntran}$), which is realized by the rotation of $R(\lambda_{LJ})$ and $R(\Omega)$. In Eq.($\ref{nyxprime}$), the LOS, $\mathbf{n}_0$,  is firstly transformed to the reference frame with  $\mathbf{k}$  aligned with  $\mathbf{L}$. And then it transforms from the frame defined by  $\mathbf{L}$ to the frame defined by the spin angular momentum, $\mathbf{S}_1$, as that in Eq.($\ref{ntran}$).


The conservation of the total angular momentum,  $\mathbf{J}$, must be held during the precession of the two bodies,
which requires that  $\mathbf{L}$ and $\mathbf{S}$ ($\mathbf{S}=\mathbf{S}_1+\mathbf{S}_2$)  be on the opposite side
of $\mathbf{J}$ at any moment~\citep{SB76,HS82}. In other words, the plane determined by the three vectors, $\mathbf{J}$, $\mathbf{S}$ and $\mathbf{L}$ rotates around the vector $\mathbf{J}$ instantaneously. In the study of  the modulation of the gravitational wave by Spin-Orbit  coupling effect on merging binaries, the orbital precession velocity satisfying   the
conservation of the total angular momentum, and the
triangle constraint have been studied extensively~\citep{Apo94,Kidder95}.

Apparently, the precession of the two spins, $\mathbf{S}_1$ and $\mathbf{S}_2$,  can change the total spin angular momentum vector,  $\mathbf{S}$, and hence varies $\lambda_{LJ}$ through the triangle formed by  $\mathbf{S}$, $\mathbf{L}$ and $\mathbf{J}$.   Then,  $\lambda_{LJ}$ can simply be obtained by, $\lambda_{LJ}=\sin^{-1}(\sqrt{S_x^2+S_y^2}/L)$, as shown in Fig.~1.

Moreover, $\Omega$ can be easily obtained by the precession phase of the two spins,  $\Omega=\tan^{-1}(-S_x/S_y)$,
where
$S_x=S_{1}\sin\lambda_{1}\cos\eta_{1}+S_{2}\sin\lambda_{2}\cos\eta_{2}$ and
$S_y=S_{1}\sin\lambda_{1}\sin\eta_{1}+S_{2}\sin\lambda_{2}\sin\eta_{2}$
in which $\eta_{1} = \dot{\Omega}_{1}t+\eta_{10}$, $\eta_{2} = \dot{\Omega}_{2}t+\eta_{20}$ (where $\eta_{10}$ and $\eta_{20}$ are initial phases).

Therefore, just starting from the very basic principles, the conservation of the total angular momentum, $\mathbf{J}$, the variation of  $\lambda_{LJ}$  and $\Omega$ required in Eq.($\ref{nyxprime}$)  can be obtained  respectively.

Putting  the precession phase of the pulsar (in the reference frame
defined by  $\mathbf{L}$), $\Psi = \eta_{1}-\Omega$,   and $\lambda_{1L}$
the misalignment angle between the spin angular momentum, $\mathbf{S}_1$, and $\mathbf{L}$,
into Eq.($\ref{nyxprime}$),  the additional time delay  can be calculated, which predicts a quasi-periodic variation. Notice that  the polar angle of $\mathbf{S}_1$ and $\mathbf{S}_2$ with respect to $\mathbf{L}$ is treated as  equal to $\lambda_{1}$ and $\lambda_{2}$, the misalignment angles with  $\mathbf{J}$ respectively, which are good approximations  in the case of $\lambda_{\rm LJ}\ll 1$.


Besides the azimuthal effect, the geodetic precession
can induce latitude effect and hence vary the pulse profile  as well, as shown in Fig.1. The impact angle between the beam axis and LOS is given,
 $\beta=\cos^{-1} n_z^{\prime}-\alpha$, from which the pulse width can be obtained~\citep{Lyne88},
\begin{equation}\label{width}
    \mathcal{W}(t) = 2\cos^{-1}[\frac{\cos \rho- \cos\alpha\cos(\alpha+\beta)}{\sin\alpha\sin(\alpha+\beta)}]
\end{equation}
Notice that the width variation of Eq.($\ref{width}$), is derived under the assumption of circular beam, the geometry of which needs only two parameters, half-opening angle of the beam, $\rho$ and magnetic inclination angle, $\alpha$, as shown in Fig.~1.

\section{Application to three pulsars}

Having simple modification to the geodetic precession induced timing effect, the new model is applied  to three pulsars\citep{Lyne10,Kramer06}.
In the fitting of timing residuals,
we focus on  whether the very basic features predicted by  the model fit the
quasi-periodic characteristics exhibited in these pulsars.

Monte Carlo method is performed  to search the best combination of parameters that
make Eq.($\ref{nyxprime}$) fit the timing noise of the three pulsars, the best fit parameters are given in Table~1.
In the fitting of a timing noise structure, such as Fig.2-4,  10 parameters are used,  three binary parameters, $P_{\rm b}^{\ast}$, $M_{\rm r}$, $I$;  and six parameters for the spin angular momenta of the two bodies, $\mathbf{S}_1$ and $\mathbf{S}_2$, which are their magnitude ($S_1$,$S_2$), misalignment angle with  the total angular momentum ($\lambda_1$,$\lambda_2$), precession phases ($\eta_1$,$\eta_2$), as well as the parameter, $\phi_{0}$, representing the initial phase of $\Theta(t)$ in Eq.($\ref{nyx}$), as shown in Table~1.

Although the timing noise displayed in the three pulsars
appear different, the relatively rapid oscillations lie on lower-frequency structures is common for all of them, which differs only in relative amplitude of oscillations and in the time scale of quasi-periods.

Among the three pulsars with significant timing noise, two pulsars,  PSR B1540-06 and PSR B1828-11 have observations both in timing residual and variation of pulse width\citep{Lyne10}.  In our model, these two effects are fitted by Eq.($\ref{nyxprime}$) and  Eq.($\ref{width}$) respectively.

\begin{figure}
\begin{center}
\includegraphics[width=0.5\textwidth]{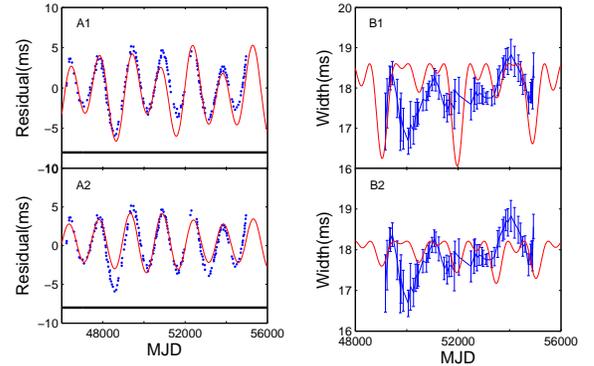}
\end{center}
\caption{\small The observed (dotted) and fitted (solid) timing residual and pulse width variation for PSR B1540-06. The upper panels, A1 and B1, show the fitted timing residual and variation of pulse width corresponding to the solution  with large mass ratio in Table~1. And the bottom panels,  A2 and B2, correspond to that of small mass ratio in Table~1.
The  horizontal line indicates that this pulsar is always at on-state.  \label{1540dtres}}
\end{figure}

As shown in Fig.($\ref{1540dtres}$) and Table~1, the fitting of both the timing residual and the variation of pulse width of  PSR B1540-06
give two solutions, the mass ratio of which are $M_r=2.63$ and $M_r=0.51$ respectively as shown in Table~1. This means companion mass of 3.7$M_{\odot}$ and 0.7$M_{\odot}$ respectively, in the case of pulsar mass of $m_1=1.4M_{\odot}$.
As shown in Table~1, an orbital period of 35 minutes and  companion mass of 3.7$M_{\odot}$   suggests  a black hole companion to the binary,
well this solution is limited by the large  projected semi-axis, $x$ predicted.  


The periodicities of 250, 500 and 1000 days have been reported in the rotation rate and pulse shape of  PSR B1828-11~\citep{Stairs00}.
These periodicities have been interpreted as being caused by free precession of this pulsar, even though it had been argued that this was not possible in the presence of the superfluid component believed to exist inside neutron stars~\citep{Lyne10}.
Apparently, the geodetic precession induced timing and profile variation avoids this difficulty automatically.


The  fitting of timing residual and pulse profile variation of
PSR B1828-11 also gives two numerical solutions. One  solution corresponds to precession periods of  $2\pi/\dot{\Omega}_1=503$d and $2\pi/\dot{\Omega}_2=592$d respectively, which accounts for the $\sim$500d period  shown in Fig.($\ref{1828dtres}$). The discrepancy between them corresponds to a period of 3000d which is responsible for the low-frequency structure displayed in Fig.($\ref{1828dtres}$). And the $\sim$250d period can be explained by the period  $2\pi/(\dot{\Omega}_1+\dot{\Omega}_2)$.


The discrepancy between the observed and predicted variation of pulse width displayed in PSR B1540-06 and PSR B1828-11 is mainly due to the assumption of a circular emission beam.


PSR B1931+24  behaves as an ordinary isolated radio pulsar for 5 to 20 days
long and switches off for the next 25 to 35 days. Such on and off
repeat quasi-periodically. Moreover, it spins down $50\%$ faster when it is on than when it is off.
These two unrelated phenomena can be interpreted by the  geodetic precession too.
As shown in Fig.($\ref{1931m}$), the timing residual of PSR B1931+24 can also be fitted by the rapid and slow oscillation of quasi-period of 7d and 45d respectively, which correspond to the precession period of the two bodies respectively. Although there are much fewer observational timing residual data points than the other two pulsars, the correlation of timing residual with the on and off states, as shown in   Fig.($\ref{1931m}$), imposes strong constraints on the fitting parameters. The phase discrepancy between timing residual and the  on and off states is likely due to the assumption of circular emission beam.

The rapid precession of the spin axis of  PSR B1931+24 suggests
that it is in a binary pulsar system with period of $\simeq 1$minute as shown in Table~1. Furthermore, the fitting of the timing residual of  PSR B1931+24, as shown in Fig.($\ref{1931m}$), automatically explains  the spin down $50\%$ faster when it is on than when it is off occurred on this pulsar.

The quasi-period of 40d  corresponds to so large a variation in misalignment angle between the LOS and the pulsar spin axis that
can make the pulsar emission beam outside the LOS. Thus explains the on-off state
and hence the intermittency exhibited in this pulsar.

Comparatively, in the  other two pulsars, the amplitude of change of $\beta$ as shown in Fig~1, is not sufficient to move the emission beam out off LOS, although the geodetic precession varies both the timing residual and pulse profile.

\section{Discussions and Conclusions}

The first numerical simulation to both  timing noise and
pulse profile variation are  performed, from which
the orbital periods of these pulsars are of 1-35 minutes. The quasi-periodicity feature of  timing noise, with rapid oscillations lying on lower frequency structure,
is well explained in the context of  the geodetic precession of an unseen binary system,
which  induces additional motion of the pulsar spin axis.

Timing noise without such a  character, e.g.,  PSR B0954+08, PSR B0642-03, PSR B1822-09 and PSR B0950+08~\citep{Lyne10}, cannot be fitted by our model. In other words, the timing noise of these pulsars may have other origins.

\begin{figure}
\begin{center}
\includegraphics[width=0.6\textwidth]{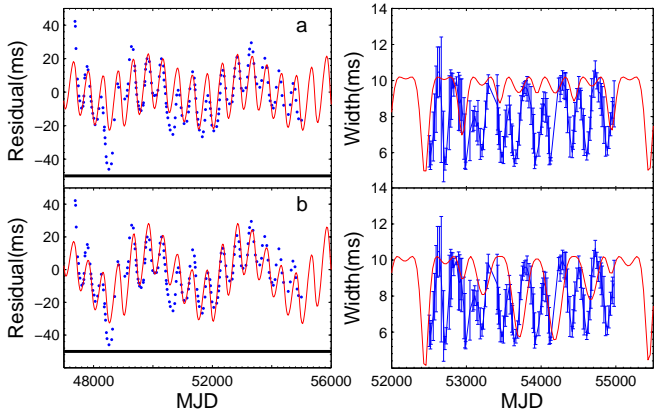}
\end{center}
\caption{\small same as Fig.(\ref{1540dtres}) but for PSR B1828-11.\label{1828dtres}}
\end{figure}

\begin{figure}
\begin{center}
\includegraphics[width=0.5\textwidth]{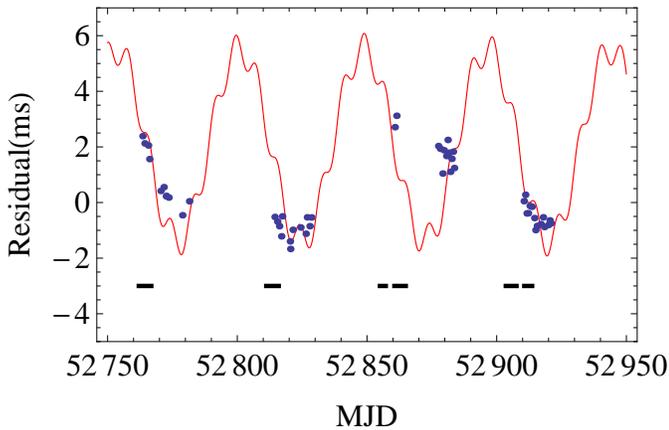}
\end{center}
\caption{\small The observed and fitted timing residual for PSR B1931+24. The dots are the observed timing residual and  the solid curve represents  the fitted one. The   horizontal lines represent the on states of the pulsar predicted by the model.\label{1931m}}
\end{figure}

As shown in Fig.2-4, it  clearly shows that the geodetic precession induced timing effect tends to be more regular, while the true timing noise appears more irregular. Such a deviation may have three reasons.
(a) The timing residual like  Fig.($\ref{1828dtres}$) is obtained by fitting the time of arrival by the rotation equation with parameters, $\nu$, and $\dot{\nu}$ (or $\ddot{\nu}$). The values of these parameters can change the shape of the timing residual. (b) Throughout the fitting, we assume circular emission beam,  while the actually beam can be non-circular, which  can also lead to irregularities  in both timing residual and pulse profile change. (c) There can be other effects like the quadruple moment in the companion star. Due to more parameters and assumptions would be introduced to account for such issues, we leave it in the future investigations.


As shown in Table~1 and the pulsar mass range discussed above, a typical companion mass of 0.25---0.34$M_{\odot}$, corresponds to  the Roche lobe radius of the companion star, $R_{2}$,
\begin{equation}
\label{R2}
R_{2}=0.462a(\frac{M_{2}}{M_{1}+M_{2}})^{1/3}\simeq3\times10^{9} (cm),
 \ \
\end{equation}
where the semi-major axis of the binary, $a$, is
$a=1.6\times10^{10}$cm (Notice that small $a_p\sim10^7 cm$ is due to the nearly face one orbital inclination and the small mass ratio, $m_2/m_1$).  The Roche lobe radius of Eq.($\ref{R2}$) is not much larger than the typical radius of White dwarf (WD) star of $\sim 1\times10^{9}$cm. Thus, the possibility of a WD as companion star cannot be excluded. Although more likely candidate would be  a strange star~\citep{Dey98}, a quark star~\citep{Xu03} or even a black hole.

The fitting parameters of Table~1 indicates that the orbital inclination angle, $i\simeq I$, influences the amplitude of the predicted timing residual. Whereas, the semi-major axis of a binary, $a$, can influence the precession velocity but not the amplitude of predicted timing residual. For a binary with a precession period of a few years, the typical semi-major axis, $a$, is of $\sim 10^{10}$cm, and with a small $i$, the projected semi-major axis, $x$, can be a few ms. If such pulsars  are observed in the single pulse mode at time scale of several or tens of the predicted orbital period as shown in Table~1, their  ultra-compact binary nature could be revealed.

The first numerical simulation to the timing residual and its correlated phenomena given by this paper provides an extremely simple scenario to the link among the timing noise,  pulse-shape variability and  intermittency. The binary assumption appears strange. However, it corresponds to acceptable companion stars and the absence of binary detection is also reasonable.
The model  can be tested by the comparison between the predicted timing residual, as shown in Fig.2-4 and the observed ones in the near future. And the binary parameters predicted in Table~1 can also be directly tested by timing series of time scale longer or much longer than the predicted orbital periods. E.g., PSR 1828-11 has  several 8-min continuous time series observed by Parkes telescope, but none of them  cover a full proposed orbital period, $\sim$20 min. A future continuous single pulse observation of time scale several orbital periods  or a few hours is expected in order to test the predicted binary nature.

This actually  proposes a new way of binary searching, which estimates the binary parameters of a candidate pulsar through fitting of its timing noise, and then test the predicted binary parameters via timing series of appropriate time scale.
The current threshold of orbital period of 90min of  radio binary pulsars could  be broken by this approach.  And  pulsar black hole binary system may be found by this method.







\begin{table*}
\begin{center}
\caption{\bf The fitting parameters for three pulsars}\label{parameters}
\resizebox{16cm}{!} {

\begin{tabular}{cccccccccccccccc}
\hline \hline

PSR & $rms$ & $\rm{P_{\rm b}^{\ast}}(0.01,2)$ & $\rm{M_{r}}(0.01,3)$  & $I(0,\pi) $ & $c_1(0.0001,0.01)$  & $c_2(0.001,0.5)$
  &  $\eta_{10}(0,2\pi)$ & $\eta_{20}(0,2\pi)$ &  $\lambda_1(0,\pi)$ & $\lambda_2(0,\pi)$ & $\phi_0(0,2\pi)$ & $\alpha(0,\pi)$ & $\rho(0,30^{\circ})$ \\

B1540--06 & $1.4$ & $0.59$ & $2.63$ & $3.10$  & $0.0018$ &
$0.026$ & $4.25$ & $3.56$ & $1.25$ & $2.51$ & $0.0052$ & $1.88$ & $7.5^{\circ}$ \\
& $1.3$ & $0.28$ & $0.51$ & $3.12$  & $0.0096$ &
$0.063$ & $0.78$ & $4.92$ & $0.59$ & $3.08$ & $-0.0036$ & $2.54$ & $4.3^{\circ}$ \\
%

B1828--11 & $13.4$ & $0.22$ & $1.21$ & $ 2.97$  & $0.0001$
& $0.08$ & $3.71$ & $3.94$ & $0.82$ & $0.94$ & $0.05$ & $2.31$ & $14.4^{\circ}$ \\
& $10.4$ & $0.19$ & $0.11$ & $3.09$  & $0.0086$ &
$0.055$ & $1.39$ & $2.05$ & $0.24$ & $1.44$ & $0.036$ & $2.90$ & $7.0^{\circ}$ \\

B1931+24 & $1.3$ & $0.016$ & $0.12$ & $ 3.12$  & $0.00010$
& $0.0057$ & $5.18$ & $1.48$ & $1.68$ & $0.87$ & $-0.016$ & $1.26$ & $10.0^{\circ}$ \\
\hline \hline
\end{tabular}
}
\end{center}
{\small The magnatic inclination angle $\alpha$ and half opening angle of emission beam, $\rho$ are obtained through fitting of pulse profile variation
 or on-off states.  All angels, except for $\rho$, are in radian.  $P_{\rm b}$ is in hour,  $M_{\rm r}\equiv m_{1}/m_{2}$; and the spin angular momenta, $c_{1}=S_{1}/L$ and $c_{2}=S_{2}/L$ are dimensionless.
$\phi_{0}$ is the initial phase for $\Theta(t)$. The range in the parenthesis followed each parameter indicates the searching
  parameters space for the fitting process. The root mean square(rms) of fitting  is given by
 $\sqrt{\sum_{j=1}^N(y^{\prime}_{j}-y_{j})^2/N^{\prime}}$, where $N^{\prime}(=N-10-1)$ is the degree of freedom, $y^{\prime}_{j}$ and $y_{j}$ are the fitted and observed timing residual,respectively. }
\end{table*}

\appendix

\section{Detailed rotation matrixes}

The four rotation matrixes are expressed as follows:
{\setlength\arraycolsep{2pt}
\begin{eqnarray}\label{rotaion}
R(\Omega)  & = &
 \left(
\begin{array}{ccc}
  \cos\Omega  & \sin\Omega & 0 \\
  -\sin\Omega & \cos\Omega & 0 \\
  0           & 0          & 1
\end{array} \right),
R(\lambda_{LJ})   =
 \left(
\begin{array}{ccc}
  1 &       0           & 0    \\
  0 & \cos\lambda_{LJ}  & \sin\lambda_{LJ} \\
  0 & -\sin\lambda_{LJ} & \cos\lambda_{LJ}
\end{array}    \right) ,  \nonumber \\
R(\Psi)  & = &
\left(
\begin{array}{ccc}
  \cos\Psi & \sin\Psi & 0 \\
  -\sin\Psi & \cos\Psi & 0 \\
  0 & 0 & 1
  \end{array} \right),
 R(-\lambda_{1})   =
 \left(
\begin{array}{ccc}
  \cos\lambda_{1} & 0 & -\sin\lambda_{1} \\
  0 & 1 & 0 \\
  \sin\lambda_{1} & 0 & \cos\lambda_{1}
  \end{array} \right). \nonumber
\end{eqnarray}
then expanding Eq.($\ref{nyxprime}$), the three components of $\mathbf{n^{\prime}}$ can be written as:
{\setlength\arraycolsep{2pt}
\begin{eqnarray}
n_x^{\prime} &=& \cos I(-\cos\lambda_{LJ}\sin\lambda_{1} +\cos\lambda_{1}\sin\Psi\sin\lambda_{LJ}) +\sin I{}\nonumber \\ && {}(\cos\Omega(\cos\lambda_{1}\cos\lambda_{LJ}\sin\Psi + \sin\lambda_{1}\sin\lambda_{LJ})+ \cos\Psi\cos\lambda_{1}\sin\Omega) \nonumber \\
n_y^{\prime} &=& \cos I\cos\Psi \sin\lambda_{LJ} + \sin I(\cos\Psi\cos\lambda_{LJ}\cos\Omega-\sin\Psi\sin\Omega) \nonumber \\
n_z^{\prime} &=& \cos I(\cos\lambda_{1}\cos\lambda_{LJ} +\sin\Psi\sin\lambda_{1}\sin\lambda_{LJ}) +\sin I{}\nonumber \\ && {} (\cos\Omega(\cos\lambda_{LJ}\sin\Psi\sin\lambda_{1}-\cos\lambda_{1}\sin\lambda_{LJ})+ \cos\Psi\sin\lambda_{1}\sin\Omega)  {} \nonumber \\ && {}
\end{eqnarray}


\begin{thebibliography}{99}



\bibitem[\protect\citeauthoryear{Apostolatos et al.} {1994}]{Apo94}
Apostolatos,T.A. , Cutler, C., Sussman, J.J.\&  Thorne,K.S. 1994,
Phys. Rev. D, 49, 6274

\bibitem[\protect\citeauthoryear{Barker $\&$ O'Connell} {1975}]{BO75}
Barker,B.M. \& O'Connell,R.F. 1975,Phys. Rev. D,12, 329

\bibitem[\protect\citeauthoryear{Cordes et al.} {1980}]{Cordes80}
 Cordes,J. M. \& Helfand, D. J. 1980,
ApJ,239, 640

\bibitem[\protect\citeauthoryear{Cordes} {1993}]{Cordes93}
Cordes,J.M.  1993, in Planets around Pulsars, Astron. Soc. Pac. Conf., ed. Philips,J.A., Thorsett,S.E. \& Kulkarni,S.R.,36, 43


\bibitem[\protect\citeauthoryear{Dey et al.}{1998}]{Dey98}
Dey, M., Bombaci, I., Dey, J., Ray, S. \& Samanta, B. C. 1998, Physics Letters B,438,123

\bibitem[\protect\citeauthoryear{Gong }{2005}]{Gong05}
Gong, B.P., 2005, Phys. Rev. Lett, 95, 261101

\bibitem[\protect\citeauthoryear{Gong }{2006}]{Gong06}
Gong, B.P., 2006, ChJAS, 2006, 6b, 273G




\bibitem[\protect\citeauthoryear{Hamilton $\&$ Sarazin} {1982}]{HS82}
Hamilton,A.J.S. \& Sarazin,G.L. 1982,
Mon. Not. R Astron. Soc.,198, 59

\bibitem[\protect\citeauthoryear{Hobbs et al. }{2010}]{Hobbs10}
Hobbs,G. , Lyne,A. G. \& Kramer, M. 2010,
Mon. Not. R. Astron. Soc.,402, 1027

\bibitem[\protect\citeauthoryear{Kidder} {1995}] {Kidder95} Kidder, L.E., 1995, Phys. Rev. D,52, 821

\bibitem[\protect\citeauthoryear{Konacki et al. }{2003}]{Konacki03}
Konacki, M. ,  Wolszczan, A.\&  Stairs,I. H. 2003,
ApJ, 589, 495

\bibitem[\protect\citeauthoryear{Kramer et al.} {2006}]{Kramer06}
Kramer,M.,Lyne, A. G., O'Brien,J. T.,Jordan,C. A. \& Lorimer, D. R. 2006,
Science, 312, 549

\bibitem[\protect\citeauthoryear{Lai et al.} {1995}]{Lai95}
Lai,D. , Bildsten,L. \& Kaspi, V.M. 1995,
ApJ, 452, 819L

\bibitem[\protect\citeauthoryear{Lyne $\&$ Manchester} {1988}]{Lyne88}
 Lyne,A. G. \& Manchester,  R. N. 1988, Mon. Not. R. Astron. Soc.,234, 477

\bibitem[\protect\citeauthoryear{Lyne et al.} {2000}]{Lyne00}
Lyne, A. G. ,  Shemar,S. L.\& Smith,  F. G. 2000,
Mon. Not. R. Astron. Soc.,315, 534

\bibitem[\protect\citeauthoryear{Lyne et al. }{2010}]{Lyne10}
 Lyne,A. G, Hobbs, G., Kramer,M., Stairs,I.\& Stappers,B. 2010,Science 329, 408

\bibitem[\protect\citeauthoryear{Manchester et al.} {2010}]{Dick10}
Manchester, R. N. et al. 2010,
ApJ, 710, 1694

\bibitem[\protect\citeauthoryear{Provencal et al.}{1998}]{Provencal98}Provencal,J. L.,Shipman,H. L., Hog, E.,\& Thejll, P.
1998,ApJ, 494, 759


\bibitem[\protect\citeauthoryear{Radhakrishnan $\&$ Cook }{1969}]{Rad69} Radhakrishnan V.\& Cook,D.J. 1969, Astrophys. Lett.,3, 226

\bibitem[\protect\citeauthoryear{Rankin et al.}{2006}]{Rankin06}  Rankin,J. M., Rodriguez, C.  \& Wright, G. A. E. 2006,Mon. Not. R. Astron. Soc.,370, 673

\bibitem[\protect\citeauthoryear{Smarr \& Blandford} {1976}]{SB76} Smarr,L.L. \& Blandford,R. 1976, ApJ,
 207, 574

\bibitem[\protect\citeauthoryear{Stairs et al.} {2000}]{Stairs00}
Stairs,I.H., Lyne, A.G. \& Shemar,S.L. 2000,Nature,406, 484

\bibitem[\protect\citeauthoryear{Shabanova}{2010}]{Sha10} Shabanova, T. V.  2000,ApJ,721, 251

\bibitem[\protect\citeauthoryear{Thorsett et al. }{2010}]{Thorsett10} Kiziltan, Bulent, Kottas, Athanasios, Thorsett, Stephen E. 2010,
 , arXiv1011.4291K

\bibitem[\protect\citeauthoryear{Wang et al.} {2007}]{Wang07}
Wang,N.,Manchester, R.N. \& Johnston, S. 2007, Mon. Not. R. Astron. Soc. 377, 1383

\bibitem[\protect\citeauthoryear{Xu}{2003}]{Xu03}
Xu,R.X. 2003, ApJ, 596, L59



\end{thebibliography}
\end{document}